\documentclass{aa}

\usepackage{graphicx,natbib}
\bibpunct{(}{)}{;}{a}{}{,}

\def\ergscm{erg~s$^{-1}$~cm$^{-2}$}

\def\cts{\hbox{{cnts/sec/pix}}}
\def\flux{erg s$^{-1}$ cm$^{-2}$}
\def\lum{erg s$^{-1}$}
\def\lummass{erg s$^{-1}$ $M^{-1}_{\odot}$}

\def\aap{A\&A}

\begin{document}

\title{Hard X-ray emission from the Galactic ridge
\thanks{Based on observations with INTEGRAL, an ESA project with
instruments and science data centre funded by ESA member states
(especially the PI countries: Denmark, France, Germany, Italy,
Switzerland, Spain), Czech Republic and Poland, and with the
participation of Russia and the USA} }

\author{Krivonos R.\inst{1,2}, Revnivtsev M.\inst{1,2}, Churazov
  E.\inst{1,2}, Sazonov S.\inst{1,2}, Grebenev S.\inst{2},
Sunyaev R.\inst{1,2}}
\institute{
              Max-Planck-Institute f\"ur Astrophysik,
              Karl-Schwarzschild-Str. 1, D-85740 Garching bei M\"unchen,
              Germany
\and
              Space Research Institute, Russian Academy of Sciences,
              Profsoyuznaya 84/32, 117997 Moscow, Russia
            }
\authorrunning{Krivonos et al.}
\abstract{We present results of a study of the Galactic ridge X-ray
emission (GRXE) in hard X-rays performed with the IBIS telescope
aboard INTEGRAL. The imaging capabilities of this coding aperture
telescope make it possible to account for the flux from bright
Galactic point sources whereas the wide field of view permits to
collect large flux from the underlying GRXE. Extensive study of the
IBIS/ISGRI detector background allowed us to construct a model that
predicts the detector count rate with $\sim1-2$\% accuracy in the
energy band 17-60 keV. The derived longitude and latitude profiles
of the ridge emission are in good agreement with the Galactic
distribution of stars obtained from infrared observations. This,
along with the measured hard X-ray spectrum of the Galactic ridge
emission strongly indicates its stellar origin.  The derived unit
stellar mass emissivity of the ridge in the energy band 17-60 keV,
$(0.9 - 1.2)\times 10^{27}$\lummass (assuming a bulge mass of $1-1.3
\times 10^{10} M_\odot$) agrees with that of local (in the Solar
neigborhood) accreting magnetic white dwarf binaries - dominant
contributors to the GRXE at these energies. In addition, the shape
of the obtained GRXE spectrum can be used to determine the average
mass of white dwarfs in such systems in the Galaxy as $\sim0.5
M_{\sun}$. The total hard X-ray luminosity of the GRXE is $L_{\rm
17-60 keV} =(3.7\pm0.2)\times10^{37}$\lum in the 17--60 keV band. At
energies 70--200 keV no additional contribution to the total
emission of the Galaxy apart from the detected point sources is
seen.
\keywords{galaxy: structure -- galaxy: bulge -- galaxy: disk -- X-rays: diffuse background -- stars: white dwarfs} } \maketitle

\section{Introduction}
Broad band studies of the radiative output of the Galaxy demonstrate that
different physical mechanisms contribute to the
brightness of the Galaxy in different energy bands. In the
near-infrared and optical spectral bands the bulk of the
emission is provided by different types of stars. In the high energy
($h\nu>$GeV) band the Galactic emission is likely a
result of interactions of cosmic rays with interstellar matter
\citep[e.g.][]{kraushaar72,kniffen78,hunter97}.

From the first all sky surveys in X-rays ($\sim$ 2-10 keV) it became
clear that in this energy band the emission of the Galaxy as a whole
is dominated by the contribution from bright point sources, mainly
accreting black holes and neutron star binaries. However, there was
also discovered emission that was not resolved into separate point
sources -- the Galactic ridge X-ray emission (GRXE, e.g.
\citealt{worrall82}). Even the significant increase of the
sensitivity of X-ray instruments over the last decades has not led
to resolving all the Galactic ridge emission into discrete sources
\citep{sugizaki01,hands04,ebisawa05}. This was considered as an
indication of a truly diffuse origin of the Galactic ridge emission.

Latest studies of the morphology and volume emissivity of the
GRXE in the energy band 3-20 keV provide convincing evidence that
the majority of the GRXE consists of a large number of stellar
type X-ray sources, namely white dwarf binaries and coronally active
stars \citep{mikej05,sazonov06}.

In particular for the energy band $>$20 keV this means that the GRXE
must be dominated by the contribution of magnetic white dwarf binaries
-- intermediate polars (IP) and polars (P).

The properties of the GRXE in hard X-rays ($>$20 keV) are not yet
well known
\citep[e.g.][]{purcell96,skibo97,kinzer99,lebrun04,terrier04,strong05}.
Assuming that the GRXE traces the stellar mass density in the Galaxy,
one can obtain a proxy of the GRXE spectrum in hard X-rays
($20-200$ keV) from the spectrum of the inner 30 pc ($\sim12^\prime$) of
our Galaxy \citep{mikej05}. However, since the the Galactic Center region
may be peculiar in many respects, study of the true GRXE in
hard X-rays is necessary.

In order to determine the origin of the hard X-ray Galactic
background it is very important to investigate whether the GRXE in hard
X-rays is distributed similar to the stellar distribution,
indicating its stellar origin, or it more closely
follows the interstellar gas density distribution, thus connecting
to the high energy gamma-ray background seen e.g. by EGRET. Does
the spectrum of the GRXE have a cutoff at energies $\sim$30-50 keV
due to the typical cutoff in spectra of magnetic CVs
\citep[e.g.][]{suleimanov05}, or it has a power law spectral shape
up to higher energies as would be expected if the Galactic
background emission were induced by cosmic ray electrons
\citep[e.g.][]{stecker77,mandrou80,sacher84,skibo93}.

Previous attempts to study the hard X-ray component of the GRXE were
severely inhibited by the poor angular resolution of the
instruments used, which precluded effective subtraction of the
contribution of bright point sources. Only now this has become
possible thanks to the hard X-ray telescopes aboard the INTEGRAL
observatory \citep{integral}. The IBIS telescope \citep{ibis} on
INTEGRAL possesses an optimal combination of properties to perform
such a study:

\begin{itemize}
\item  It has a relatively large field of
view ($\sim28^\circ\times28^\circ$ at zero response) that allows
large flux of the diffuse emission to be collected, but not too large
to preclude the construction of a GRXE map if the telescope is used as
a collimated instrument.

\item It has a possibility to detect and subtract the contribution of
  point sources

\item Its sensitivity to point sources for typical exposure times in
  the Galactic plane regions $\sim$1Ms is $\sim10^{-11}$ \flux, which for the
Galactic Center distance corresponds to a luminosity $\sim10^{35}$
\lum. Subtraction of sources with luminosities higher than this
limit allows one to avoid significant contamination of the GRXE by
point sources \cite[see e.g.][]{sazonov06}
\end{itemize}

Since its launch in 2002 INTEGRAL/IBIS has collected a large amount of
observational data on different sky regions, and in particular on the
inner Galactic plane where most of the GRXE is located.

In this work we will study the spectral and morphological properties of
the GRXE in the hard X-ray energy band 17-200~keV. At higher energies the
positronium annihilation continuum of the Galactic Center
\citep[e.g.][]{leventhal78,gehrels91,churazov05,knodlseder2005} should be carefully taken into account, which requires a different approach from the one
used in this work (which is especially related to the detector
background modeling). For this reason we leave the study of the
Galactic background emission at energies higher then $200$~keV for a
separate paper.

\section{Data set and data filtering}
\label{section:data}

For our analysis we used all the IBIS data available to us,
including public data, some proprietary data (Galactic Center
observations and Crux Spiral arm deep exposure observations) and
data available to us through the INTEGRAL Science Working Team. In
total we analyzed $\sim$33 Msec of the data (deadtime corrected
value of exposure). We considered only the data of the ISGRI
detector of the IBIS telescope, which provides data in the energy
band $\sim17-1000$ keV with high sensitivity in hard X-rays (17-200
keV) and has sufficient angular resolution ($\sim12^\prime$) for
studying crowded fields like the Galactic Center region.

The method of the sky reconstruction employed in the IBIS telescope
(coded mask imaging) does not allow one to study directly
diffuse structures that are significantly larger that the size of
the mask pixels. Therefore, in order to study large scale structures,
such as the GRXE ($\sim 100^\circ \times5^\circ$) , we should use
IBIS/ISGRI as a collimated instrument. The detector collects
photons from point sources and diffuse emission. Measurement of the point
sources contribution to the total detector count rate makes it
possible to recover the flux of the GRXE. The success of such approach
strongly depends on the accuracy of the instrumental background
modelling.

Prior to subsequent analysis we screened the data. If an individual
observation (SCW - "science window") did not fulfill all the
imposed criteria it was dropped. However, we should note that
screening of individual events is likely more flexible and could have saved
slightly more data.

We screened all the data near the beginning and end of revolutions
(due to increased background near the radiation belts), the data when
ISGRI was operated not in its main regime (modes 41 and 43), and science
windows with exposure times less then $\sim700$~s. We also applied
filtering using information about the electron count rate provided
by IREM (INTEGRAL Radiation Environment Monitor, \citealt{hajdas03}).
Analysis of the detector 10-s binned lightcurve in the energy band
17-200 keV was used to screen science windows with all types of
bursts. As a final step in the screening procedure, we filtered out all
observations which had a high level of noise on source-free images (sky
images with removed point sources).

Upon the data filtering only $\sim60\%$ of observations
were accepted for further analysis.


\begin{figure}
\includegraphics[width=0.5\textwidth]{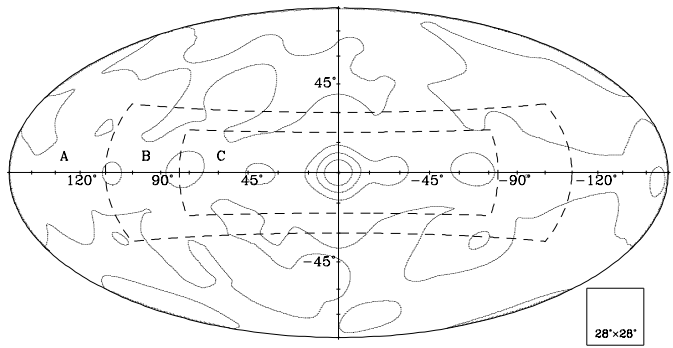}
\caption{Labeling of the sky areas used in this work. Area ``A''
($|l|>120^{\circ } \cap |b|>30^{\circ }$, excluding a
  $30^\circ$-radius circle around the Crab nebula) and ``B''
  ($90^{\circ}<|l|<120^{\circ} \cap
20^{\circ}<|b|<30^{\circ}$) contain extragalactic fields used for
calibration of the background model. Area ``B' was also used to
estimate the accuracy of the method. The Galactic ridge X-ray emission was
studied in field ``C'' observations ($|l|<80^{\circ}\cap
|b|<20^{\circ}$). Contours of the total (deadtime corrected) exposure
time are shown overimposed. The contour levels correspond to 5.3, 3.5,
1.8 Ms, and 50 ks. The IBIS FOV is shown for comparision at the right
bottom corner of the plot.} \label{fig:map}
\end{figure}

\begin{figure}
 \includegraphics[width=0.5\textwidth]{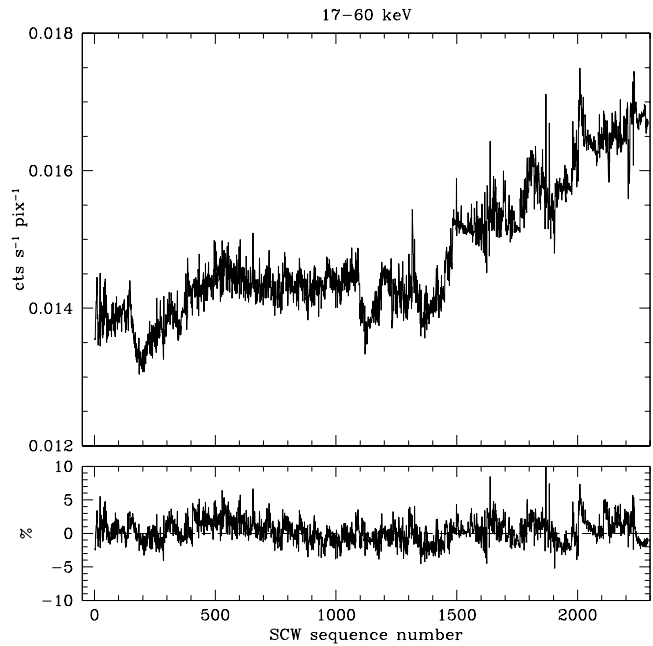}
\caption{ {\it Top panel:}  ISGRI count rate in the $17-60$~keV energy
band as a function of SCW sequence number. The pointings were
selected in area ``B'' (see Fig.\ref{fig:map}). {\it Bottom panel:}
Residuals after subtracting the model-predicted count rate from the
observed count rate (in per cent with respect to the observed count
rate).}\label{fig:model}
\end{figure}

\begin{figure}
\includegraphics[width=0.5\textwidth]{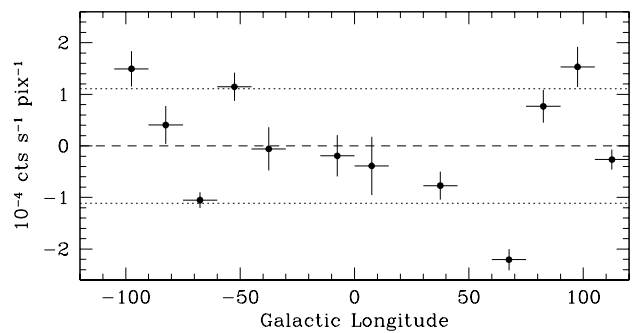}
\caption{Residuals of the detector count rate of field "B" observations
from the background model averaged over Galactic longitude. Dotted
lines represent a $1\sigma$ deviation ($1.1\times
10^{-4}$~cts~s$^{-1}$~pix$^{-1}$) of averaged values from zero.}
\label{fig:uncert}
\end{figure}

\section{Detector modelling}
\label{section:model}

\subsection{Detector background}
\label{section:model:definition}

At any given time the detector count rate of IBIS/ISGRI consists of:

\begin{itemize}
\item Cosmic X-ray Background (CXB)
\item Emission from point sources
\item Galactic ridge X-ray emission, if the field of view of the
telescope is directed towards the Galactic plane
\item Detector internal background, caused by different
processes including activation of different elements of the
spacecraft, interaction of the detector material with cosmic-rays,
etc.
\citep[e.g.][]{bloser02,terrier03}
\end{itemize}

The Cosmic X-ray Background contributes $\sim$0.5 Crab in the energy
band 17-100 keV and can be considered very uniform over the sky.
Taking into account the IBIS/ISGRI field of view, we can expect that
CXB flux variability over the sky for ISGRI will not exceed
$\sim1$\% (e.g. Sunyaev et al. 2006, in preparation). Therefore, the
contribution of the CXB to the detector count rate can be considered
as independent of observation orientation and can be estimated from
observations at high Galactic latitudes. The contribution of point
sources to the detector count rate can be almost perfectly predicted
using the telescope coded mask imaging technique (more on this
below). The list of detected sources used in our subtraction
procedure includes more than 360 sources on the entire sky.
Typically the detection limit for regions near the Galactic plane is
at the level of $\sim1$ mCrab. The complete list of sources will be
presented elsewhere.

To predict the detector count rate not caused by photons arriving from
the sky one should use a background model.

We followed two different approaches to study the ridge
morphology and its energy spectrum. For the study of the energy spectrum
we used only specially performed INTEGRAL observations for which the
systematical uncertainties are minimal (see description below, model
2), whereas for the study of the ridge emission distribution in the
Galaxy we used all the available observations.

{\it Model 1: Background model using tracers}

Among possible tracer candidates that might be used in the
background modeling we considered the IREM count rates, SPI
saturated events count rates, ISGRI veto count rates, and ISGRI high
energy bands count rates. We finally chose the ISGRI detector
count rate in the energy band $600-1000$~keV. At these energies the
effective area of the ISGRI detector is very small ($<40$ cm$^2$) and
the detector count rate is expected to be dominated
by the internal detector background.

We also took into account gain variations of the ISGRI detector, which shift
the detector background spectrum along the energy axis causing
additional variations of the count rates in the studied energy
bands.  As we are mainly interested in the energies 17-200 keV, we
estimated the ISGRI detector gain by observing the position
of strong background lines at energies $\sim$~60 keV
\cite[e.g.][]{terrier04}.

Our final model of the ISGRI background consists of a linear
combination of the $600-1000$~keV detector count rate $H$ and the gain
parameter $G$. To make allowance for possible long-term
variations of the ISGRI detector background we also included time
($T$) in the cubic polynomial form:

\begin{equation}
S=a_0 + a_1 H + a_2 G + a_3 T + a_4 T^2 + a_5 T^3
\label{eq:bgdmodel}
\end{equation}

The coefficients in eq. \ref{eq:bgdmodel} for the 17-60~keV energy band
were calculated using observations pointed away from the Galactic
Center and away from the inner Galactic plane, where the GRXE is
negligible \citep{mikej05} -- regions ``A'' and ``B'' in
Fig.\ref{fig:map} (hereafter we consider only the detector
count rates after removal of the contribution of point sources).

\begin{figure*}[t]
\hbox{
\includegraphics[width=\columnwidth,clip]{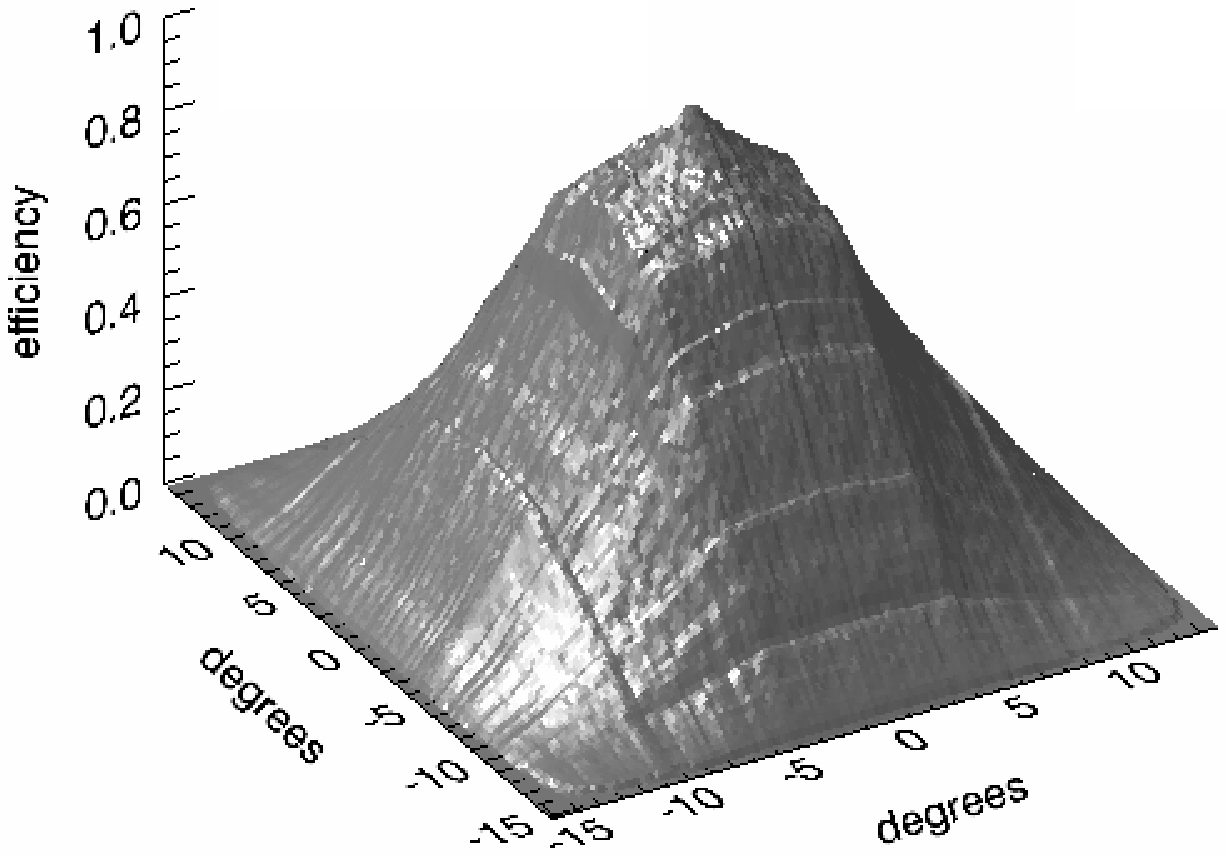}
\includegraphics[width=\columnwidth,clip]{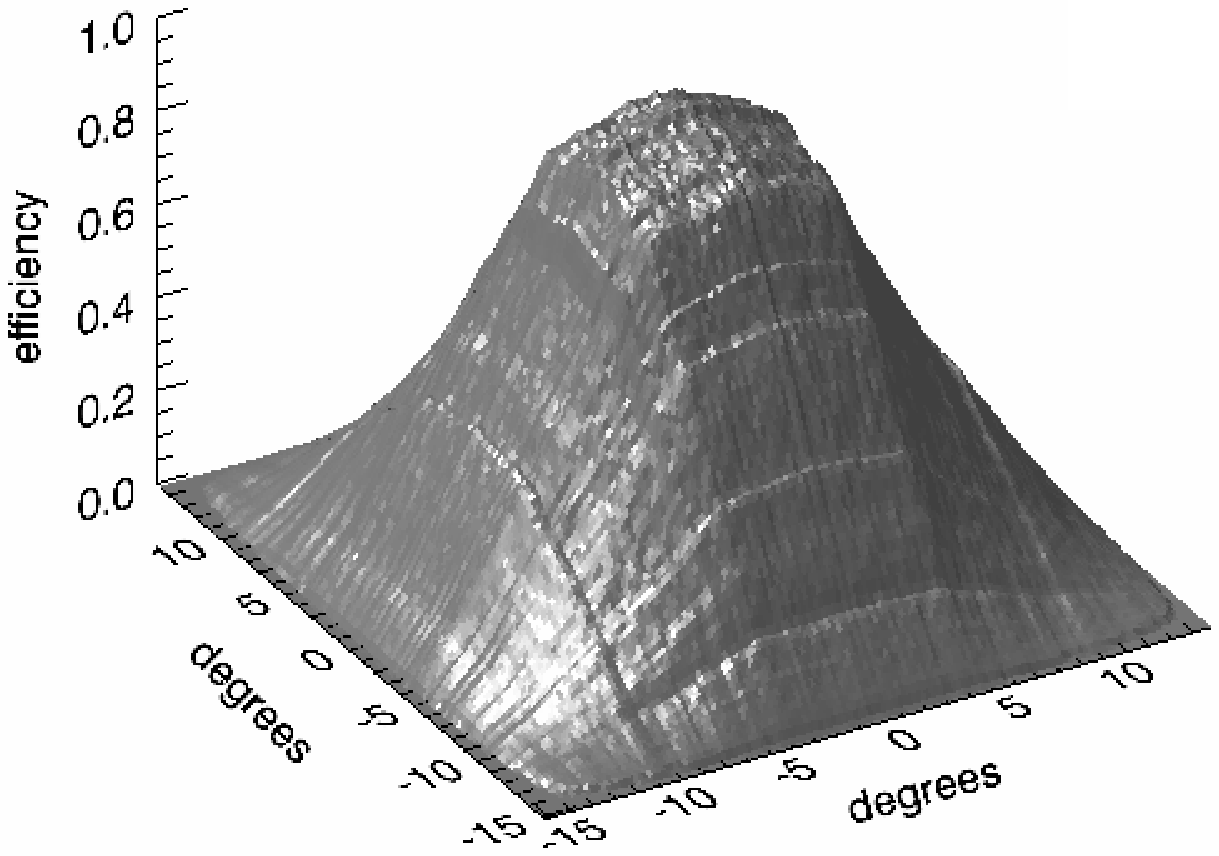}
}

\caption{IBIS/ISGRI efficiency as a function of point source
position within the field of view. The model of the IBIS/ISGRI efficiency
in the energy band 17-60 keV (left panel) and 86-129 keV (right
panel). Observations of the Crab nebula were used for the calibration.
}\label{fig:crab}
\end{figure*}

In order to test the background model we derived coefficients for
our model using pointings in region "A" Fig.\ref{fig:map} and
applied the model to pointings from region ``B''. The result of the
background model implementation is presented in Fig.\ref{fig:model}
(bottom panel). The employed background model leaves residuals with an
rms scatter $\sim1.8$\% of the mean background level
(see bottom panel in Fig.\ref{fig:model}). However, it should be
noted that this rms scatter contains significant contribution from
the pure statistical variation of the (point sources removed) detector
count rate. In order to reduce the contribution of this statistical
scatter and to reveal only systematic uncertainties we averaged the
measured field ``B'' residuals (differences between the measured
detector count rates and model predictions) over the Galactic
latitude. The resulting residuals are presented in Fig.\ref{fig:uncert}.
The root-mean-squared value of the residuals is $1.1\times10^{-4}$
\cts\ in the energy band 17-60 keV ($\sim1\%$ of the detector
background), which approximately corresponds to a flux
$\sim10$~mCrab for a Crab-like spectrum.

{\it Model 2: Rocking mode}

 Part of the Galactic Center region observations
-- Galactic Center Latitude scans (March 2005 -- March 2006) -- were
taken by INTEGRAL using a specially designed pattern, which presents
considerable advantages from the point of view of the study of the
Galactic ridge energy spectrum. The pointing direction of INTEGRAL
instruments were moved across the Galactic Center region on a time
scale $\sim10$ hours, which is smaller than that of significant
changes of the INTEGRAL/IBIS/ISGRI instrumental background. This
mode of observations turns the IBIS/ISGRI instrument into some kind
of a rocking collimator experiment. Therefore, the prediction of the
ISGRI/IBIS instrumental background in this mode of observations was
calculated using interpolation between ISGRI flux measurements done
at high ($|l|>20^\circ$) Galactic latitudes, where the surface
brightness of the Galactic background emission is negligible
\cite[e.g.][]{mikej05}

We checked the quality of the ISGRI instrumental background
subtraction on high energy channels ($>600$ keV) where the
instrumental background totally dominates. We found that
the systematical uncertainties of the background subtraction using the
employed technique ($\sim0.5\%$ of the detector count rate) do not
exceed the statistical uncertainties of the $\sim$~1 Msec dataset used. In
particular, for energy channels $\sim$100-200 keV this means
approximately 7 times better quality of the background subtraction
than when using the method described in previous paragraphs (model 1) --
$\sim$15 and $\sim$100 mCrab correspondingly.

Thus for the construction of the Galactic ridge energy spectrum we
used only these observations. Unfortunately this method cannot be
used for studying the whole Galaxy because the special pattern of
observations (Galactic latitude scans) is available only for the Galactic
Center region.

\subsection{Accuracy of the sources subtraction, IBIS collimator efficiency
 and influence of the IBIS mask transparency}

One of the key steps in studying the GRXE with the large field of
view instrument IBIS/ISGRI is the subtraction of the contribution of
bright point sources from the detector count rate.  The unsubtracted
count rate due to bright point sources can significantly distort the
map of the GRXE and its energy spectrum.

There are two main constituents of this problem: imperfect mask shadow
modeling in the software, which will result in leftover unsubtracted
count rate on the detector, and the finite opacity of the IBIS mask,
which will lead to an underestimation of source count rates by the
coded mask technique and therefore to an unsubtracted source count
rate on the detector.

In order to check the quality of our source removal procedure and
the accuracy  of our constructed IBIS collimator response function
we studied a number of observations of the Crab nebula.

Our model of the IBIS/ISGRI collimator response function includes a
geometry of the instrument and also an angle and energy dependent
absorption caused by the "NOMEX" structure, supporting IBIS mask.
(see e.g. \citealt{reglero01}). Examples of used response functions
in the energy bands 17-60 keV and 86-129 keV are presented in
Fig.\ref{fig:crab}.

The residuals between the predicted count rates during Crab
observations (which include the results of our background model and
our collimator response function model) and the actually measured
detector count rates do not demonstrate any dependence on
source flux and the rms scatter of the residuals does not exceed the
uncertainty of the our detector background model estimated in
section \ref{section:model:definition}. Therefore, we can conclude
that neither our source removal procedure nor finite mask opacity in
the energy range 17-200 keV introduced additional systematic
uncertainties into our analysis.



\section{Results}

\begin{figure}[htb]
\includegraphics[width=0.5\textwidth]{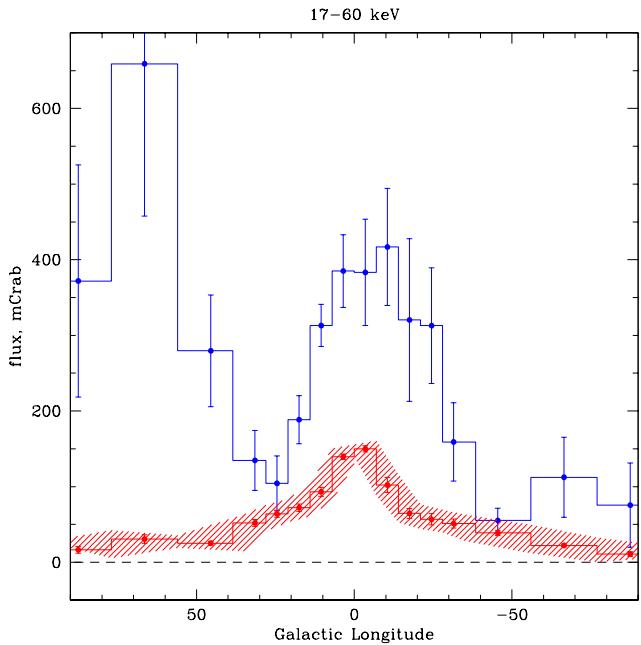}
\caption{Longitude profile of the ridge emission in the $17-60$~keV
energy range. Only observations during which the center of the ISGRI
field of view was directed within $|b|<5^{\circ}$ were used for the
construction of this profile. The profile is shown by the red
histogram with the shaded region representing systematic
uncertainties. The contribution of source emission is shown by the
blue histogram. Error bars represent rms-deviations of individual
measurements of the summed point source fluxes from the average flux
values in bins.}\label{fig:long}
\end{figure}

\begin{figure}[htb]
\includegraphics[width=0.5\textwidth]{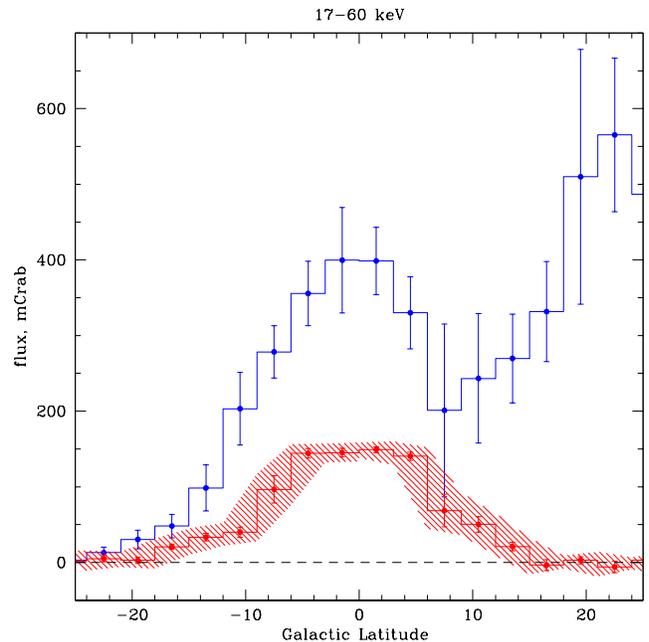}
\caption{
Ridge emission latitude profile ($|l|<5^{\circ}$). The plot
description is the same as for Fig.\ref{fig:long}.}\label{fig:lat}
\end{figure}

Using the method described above for each INTEGRAL/IBIS/ISGRI
observation we obtain two numbers in any considered energy channel:
1) the summed detector count rate caused by resolved point sources
and 2) the detector count rate left after subtraction of the
contribution of point sources and modeled detector background.

The possible remaining contribution of undetected point sources on the
detector can be estimated using the luminosity function of Galactic
X-ray sources \citep{grimm02,sazonov06}. The majority of the inner
Galactic plane was observed by INTEGRAL/IBIS for more than
0.5-0.8 Ms. Such an exposure corresponds to an IBIS/ISGRI
detection sensitivity $\sim1$ mCrab $\sim10^{-11}$ \flux\ in the
energy band 17--60 keV, which in turn corresponds to a source
luminosity $\sim10^{35}$ \lum\ for the Galactic Center distance.
The contribution of sources brighter than this limit was subtracted
from the detector count rate. Therefore, from Fig.12 of
\cite{sazonov06} we can conclude that the contribution of undetected
point sources does not significantly affect the emission of the
GRXE.

\subsection{Morphology}
\label{section:morphology}

We constructed longitude and latitude profiles of the integrated
emission from point sources and of the hard X-ray Galactic ridge
emission. For construction of the longitude profile of the GRXE we
selected those INTEGRAL observations where the IBIS axis was
directed within $5^\circ$ of the Galactic plane and then averaged
the obtained GRXE flux measurements in nearby longitude bins. The
obtained profile is presented in Fig.\ref{fig:long}. The blue
(upper) histogram represents the summed flux of point sources seen
by the IBIS/ISGRI detector. Error bars on this histogram represent
the rms deviations of the individual measurements of the summed
point source fluxes. On this plot one can clearly see the
contributions of a number of sources in the Galactic Center region
($l\sim0^\circ$) and well-known bright Galactic X-ray sources Cyg
X-1 ($l\sim70^\circ$), GRS 1915+105 ($l\sim45^\circ$), and GX 301-2
($l\sim-60^\circ$).

The latitude profiles (detected sources and unresolved background)
were obtained by averaging those GRXE flux measurements made when the
IBIS telescope was directed within $|l|<5^\circ$
(Fig.\ref{fig:lat}). The contributions of bright Galactic Center sources
and Sco X-1 ($b\sim20^\circ$) are clearly seen.

As we used the IBIS/ISGRI with the field of view
$\sim15^\circ\times15^\circ$ (FWHM) as a collimated instrument, the angular
resolution of our resulting profiles is approximately $\sim15^\circ$.
This angular resolution is not ideal for studying the innermost
(angular scales 1-2$^\circ$) regions of the Galaxy and latitude
profiles of the GRXE (exponential scale heights 1.5-3$^\circ$, see
e.g. Revnivtsev et al. 2005). Taking into account also the limited
accuracy of our obtained IBIS collimator response function we
concluded that at the present level of uncertainties involved in our
analysis we cannot make an accurate multiparameter model fitting of
the GRXE volume emissivity.

In order to extract information about the three-dimensional
structure of the Galactic ridge in hard X-rays we have compared the
profiles of GRXE proxies (with known properties) with those obtained
by us from INTEGRAL/ISGRI data. In particular, the current
understanding of the GRXE morphology implies that the best tracer of
the GRXE is the near infrared surface brightness \citep{mikej05}.

The map of the Galaxy in the near infrared spectral band was obtained
using data of COBE/DIRBE observations (zodi-subtracted mission average
map provided by the LAMBDA archive of the Goddard Space Flight Center,
http://lambda.gsfc.nasa.gov). In order to reduce the influence of the
interstellar reddening we considered DIRBE spectral band $4.9\mu$m.

We applied first-order corrections to the NIR map of the Galaxy
obtained by COBE/DIRBE.  We assumed that the intrinsic NIR color
temperature (i.e. the ratio of intrinsic surface brightnesses $I_{1.2\mu
m}$ and $I_{4.9\mu m} $) of the Galactic disk and the Galactic
bulge/bar is uniform and its true value can be derived at high
Galactic latitudes where the interstellar reddening is negligible.
Then the foreground extinction map may be expressed as:
\[
A_{\rm 4.9\mu m}={-2.5\over{A_{1.2\mu m}/A_{4.9\mu
m}-1}}\left[\ln\left({I_{1.2\mu m}\over{I_{4.9\mu
m}}}\right)-\ln\left({I_{1.2\mu m}^{0}\over{I_{4.9\mu
m}^{0}}}\right)\right]
\]

Here the $A$ values are the reddening coefficients at different
wavelengths. We have used the interstellar reddening values from works
of \cite{lutz96,indebetouw05}. The employed correction of course
removed only main effects of interstellar extinction on the COBE/DIRBE
map, therefore we do not expect that the obtained COBE/DIRBE map and
profiles have accuracy higher than $\sim10\%$.

\begin{figure}[htb]
\includegraphics[width=0.5\textwidth]{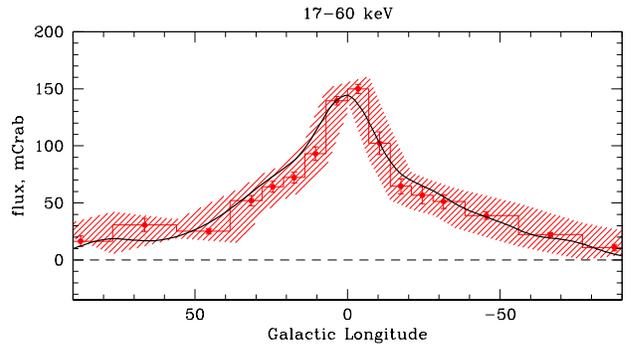}
\caption{Longitude profile of the GRXE measured by
INTEGRAL/IBIS/ISGRI (histogram and shaded region) in the 17-60 keV
energy band along with the intensity profile of the Galactic NIR
emission obtained by COBE/DIRBE at 4.9 $\mu$m (solid line). The NIR
map was convolved with the IBIS collimator response. Normalization of
the NIR profile is determined from X-ray-NIR correlation function (see
Fig.\ref{fig:corr})}\label{fig:lon_cobe}
\end{figure}

\begin{figure}[htb]
\includegraphics[width=0.5\textwidth]{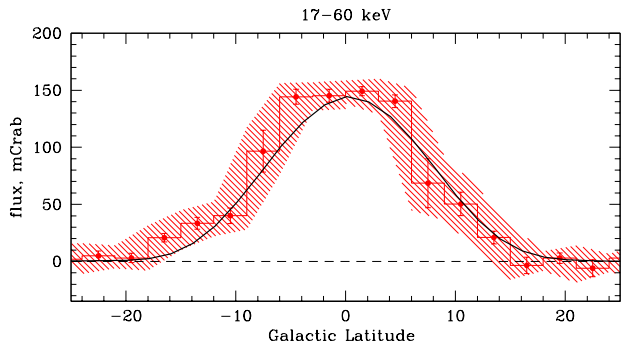}
\caption{
Latitude profile of the GRXE. IBIS telescope pointings were selected
within $|l|<5^{\circ}$. For plot description see
Fig.\ref{fig:lon_cobe}.  }\label{fig:lat_cobe}
\end{figure}

\begin{figure}[htb]
\includegraphics[width=0.5\textwidth]{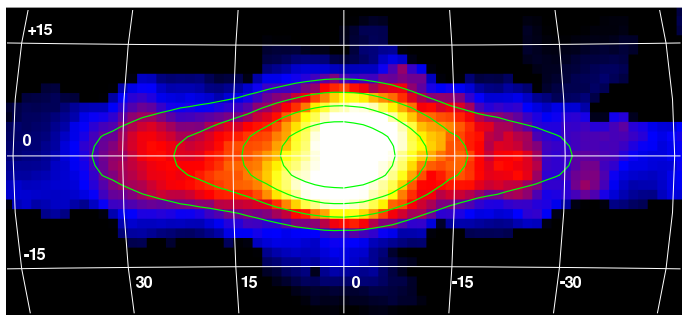}
\caption{
Map of the Galactic diffuse emission observed by INTEGRAL/IBIS/ISGRI
in the energy band $17-60$~keV. The map was convolved with a gaussian filter
($\sigma=1.3^{\circ}$). Contours represent the near infrared intensity measured
by COBE/DIRBE at $4.9\mu m$. NIR contours were convolved with the IBIS
collimator response function. The contour levels correspond to
$1.0,1.4,1.8,2.2\times10^{-5}$ \ergscm per IBIS FOV.}\label{fig:ridgemap}


\end{figure}

\begin{figure}[htb]
\includegraphics[width=\columnwidth]{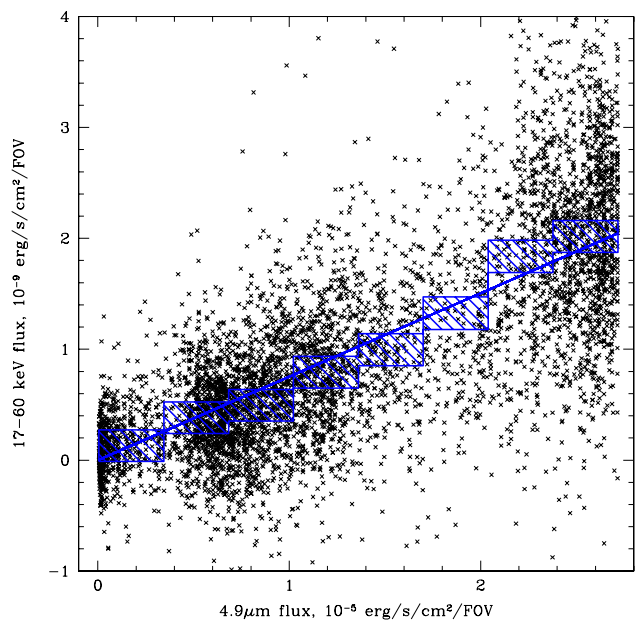}
\caption{Correlation of the near infrared measurements of COBE/DIRBE
at 4.9 $\mu$m with the hard X-ray fluxes observed by IBIS/ISGRI in
the energy band 17-60 keV. Every point corresponds to a single
INTEGRAL/IBIS pointing. Value of axis $y$ is the GRXE flux measured
by IBIS/ISGRI when pointed at a certain sky position, value of axis
$x$ is the 4.9$\mu$m flux obtained by convolution of the DIRBE NIR
map with the IBIS/ISGRI collimator response function. Scatter of
points in "y" direction is compatible with statistical and
systematic uncertainties of IBIS/ISGRI measurements. Values
obtained by averaging of hard X-ray measurements in NIR flux bins
are shown by blue squares. The half height of the squares represents
the uncertainty of our measurements. In most cases the uncertainty
is dominated by systematic errors of our background model
$\sim10$~mCrab. The solid blue line is the linear correlation of NIR and
hard X-ray fluxes: $F_{\rm 17-60 keV}/F_{\rm 4.9\mu
m}=(7.52\pm0.33)\times 10^{-5}$. }\label{fig:corr}
\end{figure}

\begin{figure}[htb]
\includegraphics[width=0.5\textwidth]{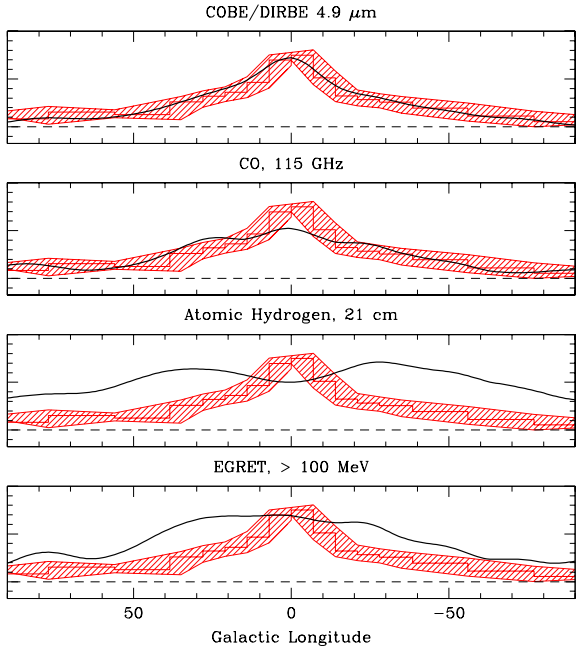}
\caption{Profile of the GRXE in hard X-rays (17-60 keV) observed by
IBIS/ISGRI along with the profiles of EGRET gamma--ray background,
neutral hydrogen (HI) emission, and molecular gas (CO) emission. All
the profiles were convolved with the IBIS/ISGRI collimator response
function and arbitrary normalized for better
visibility.}\label{fig:morpho}
\end{figure}

\begin{figure}[htb]
\includegraphics[width=\columnwidth]{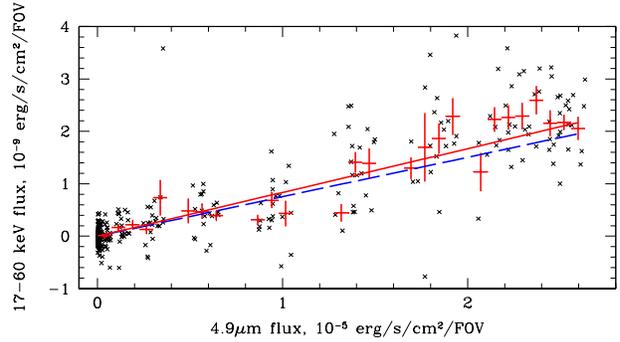}
\caption{
Correlation between NIR (COBE/DIRBE, 4.9 $\mu$m) and hard X-ray
(IBIS/ISGRI, 17-60 keV) fluxes. ISGRI detector count rate was measured
using Galactic Center Latitude scans in ``rocking mode'' approach
(background model ``2'', see text). Each black point represents individual
measurement. ISGRI detector count rate was averaged over NIR flux as
shown by red data points. Linear correlation coefficient was found as
$(8.44\pm0.28)\times 10^{-5}$ (red line). Blue dashed line represents NIR to
hard X-ray correlation obtained using all available observations
(background model ``1'').}
\label{fig:corr_interpol}
\end{figure}

\begin{figure}[htb]
\includegraphics[width=\columnwidth]{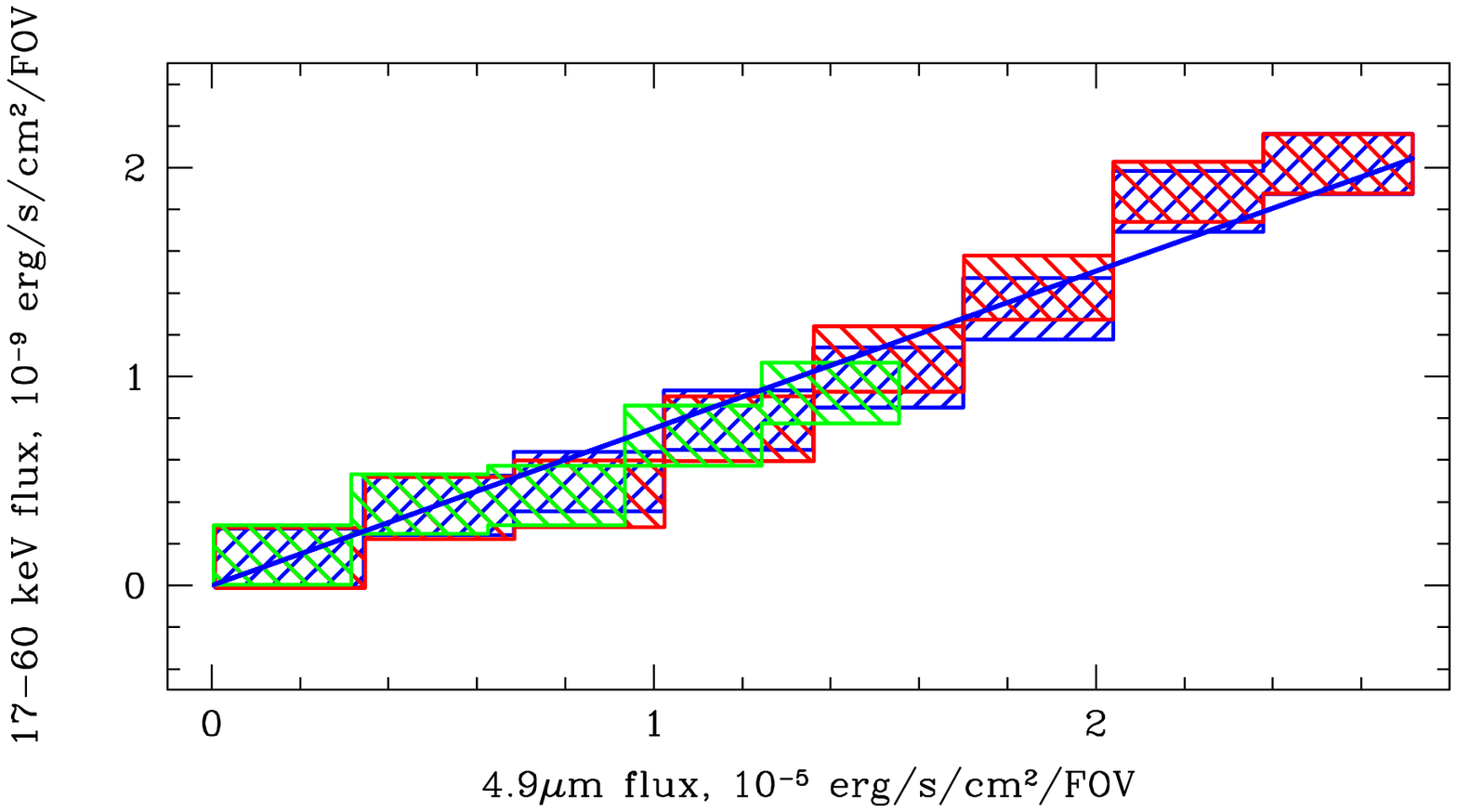}
\caption{NIR- and hard X-ray correlation (similar to
Fig.\ref{fig:corr}) measured in the Galactic bulge
($|l|<10^{\circ}$, red boxes) and Galactic disk ($|l|>20^{\circ}$,
green boxes) regions separately. The linear correlation between NIR
and hard X-ray fluxes measured using all GRXE observations is shown
in blue (shaded region and fit).} \label{fig:corr_bulge_disk}
\end{figure}

The map of the NIR intensity was then convolved with the IBIS/ISGRI
collimator response function (see Fig.\ref{fig:crab}). The resulting
longitude and latitude profiles of the COBE/DIRBE NIR intensity are
shown by the solid line on Fig.\ref{fig:lon_cobe} and \ref{fig:lat_cobe}
correspondingly. We also constructed a map of the IBIS/ISGRI
surface brightness distribution of the GRXE in the $17-60$~keV energy
band (see Fig.\ref{fig:ridgemap}). The map of the NIR intensity is shown
by contours. Correlation of observed values of hard X-ray flux with the
NIR fluxes is presented in Fig.\ref{fig:corr}.

It is clearly seen that the GRXE intensity distribution very closely
follows the NIR intensity distribution and thus traces the stellar mass
density in the Galaxy. In order to show that the correlation of the
hard X-ray GRXE with the cosmic--ray induced gamma--ray background
emission is not nearly as good as its correlation with the NIR
intensity, we present Fig.\ref{fig:morpho}. Here one can see the
distributions (convolved with the IBIS/ISGRI collimator response
function) of the EGRET gamma--ray background, Galactic neutral hydrogen
(HI), and molecular gas (CO emission).

We can conclude that the emissivity profile of the GRXE in hard
X-rays (17-60 keV) supports the finding of \cite{mikej05} that the GRXE
traces the stellar mass distribution. This allows us to estimate the
emissivity of the GRXE in hard X-rays using the known NIR luminosity
measured with COBE/DIRBE observations.

The ratio of NIR- and hard X-ray intensities averaged over the whole
Galaxy is $F_{\rm 17-60 keV}/F_{\rm 4.9\mu m}=(7.52 \pm 0.33)\times
10^{-5}$ (Fig.\ref{fig:corr}). The ratio derived using data of
Galactic Center Latitude scans only (background model 2, see Sec.
\ref{section:model}) is $F_{\rm 17-60 keV}/F_{\rm 4.9\mu
m}=(8.44\pm0.28)\times 10^{-5}$ (see Fig. \ref{fig:corr_interpol})
agrees with that measured over the whole Galaxy.

The ratios averaged over the Galactic bulge ($|l|<10^{\circ}$) and
the Galactic disk ($|l|>20^{\circ}$) regions separately are
$(7.73\pm0.34)\times 10^{-5}$ and $(6.53\pm0.72)\times 10^{-5}$
correspondingly (Fig.\ref{fig:corr_bulge_disk}). Note that there is
no statistically significant difference in the obtained ratios for
the bulge and disk regions.

Using the $4.9\mu m$ luminosity of the Galactic bulge $L_{4.9\mu
m}^{\rm bulge}=4.3\times10^{7}L_{\odot}$ by \citealt{dwek95} and
NIR-to hard X-ray flux ratio we can estimate the 17-60 keV
luminosity of the Galactic bulge as $(1.23\pm0.05)\times10^{37}
$\lum. Assuming a Galactic bulge mass of $M_{\rm bulge}=1-1.3 \times
10^{10} M_\odot$ (e.g. \citealt{dwek95}), we can estimate the unit
stellar mass hard X-ray emissivity of the GRXE as $L_{\rm 17-60
keV}/M_{\rm bulge}=0.9-1.2 \times 10^{27}$ erg/s/$M_\odot$. Taking
the disk-to-bulge mass ratio to be $\sim2$, we can estimate that the
total hard X-ray (17-60 keV) luminosity of the Galaxy in the ridge
emission is $(3.7\pm0.2)\times10^{37}$ \lum.

\begin{figure}[htb]
\includegraphics[width=0.5\textwidth,bb=30 170 600 730,clip]{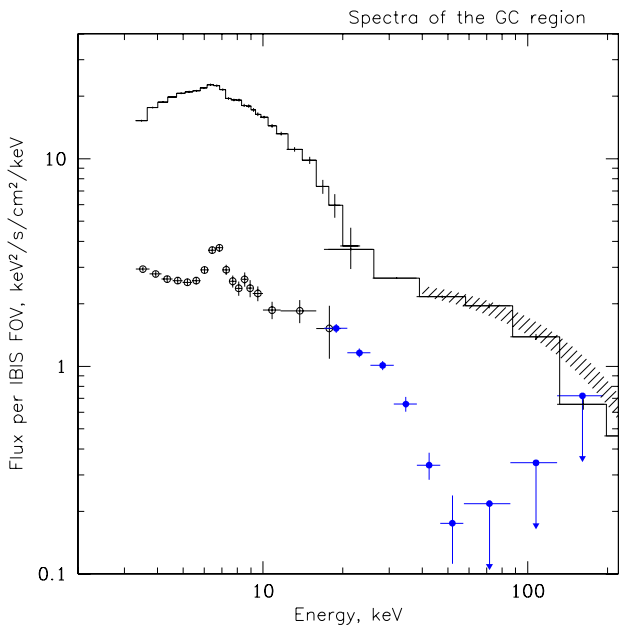}
\caption{Spectrum of the Galactic Center region in the energy band
3-200 keV as it would be seen by a $\sim15^\circ\times15^\circ$
field-of-view instrument. The data points at energies $>17$~keV are
obtained by IBIS/ISGRI. The arrows at energies $>$60 keV are $2\sigma $--upper
limits. The points at $3-20$ keV were obtained from the
data of RXTE/PCA (Galactic bulge scan data taken on March 15, 1999)
and scaled to match the IBIS/ISGRI points at $\sim20$ keV. Circles
show the spectrum of the GRXE, open circles - RXTE/PCA data, and filled
circles - INTEGRAL/IBIS/ISGRI data. The histogram shows the integrated
emission of detected point sources. The shaded region represents a model
fit to the spectrum of the GC region measured by OSSE. The contribution
of positron annihilation radiation, which consists of a 511 keV line
and annihilation continuum, was subtracted \citep{kinzer99}. Its
normalization was scaled to match that of the IBIS/ISGRI points at
30-40 keV }\label{fig:gcspec}
\end{figure}

\subsection{Spectrum}
Using the data from Galactic latitude scans, which have the smallest
systematic uncertainties of the ISGRI background subtraction, we
obtained the spectrum of the GRXE in the energy band 17-200 keV. It is
presented in Fig.\ref{fig:gcspec} together with the summed spectrum of
all detected point sources in the Galactic Center region.

\begin{figure}
  \includegraphics[width=\columnwidth]{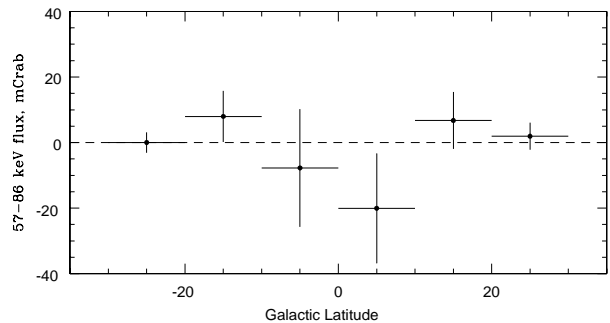}\\
  \caption{Latitude profile of the GRXE in the energy band $57-86$~keV. Only
  an upper limit on the ridge emission in this energy band can be obtained}\label{fig:lat_highen}
\end{figure}

Note that after the subtraction of bright point sources detected by
IBIS/ISGRI we do not detect any additional hard X-ray emission at
energies $\sim$60-200 keV. Our $2\sigma$ upper limit on such
emission is $\sim60$~mCrab for the IBIS field of view and for the energy
band $57-86$~keV. In order to demonstrate that the GRXE above
$60$~keV vanishes we present the latitude profile of the IBIS/ISGRI
detector count rate in the energy band 57-86 keV (Fig.
\ref{fig:lat_highen}).

\begin{figure}[htb]
\includegraphics[width=\columnwidth,bb=30 150 600 730,clip]{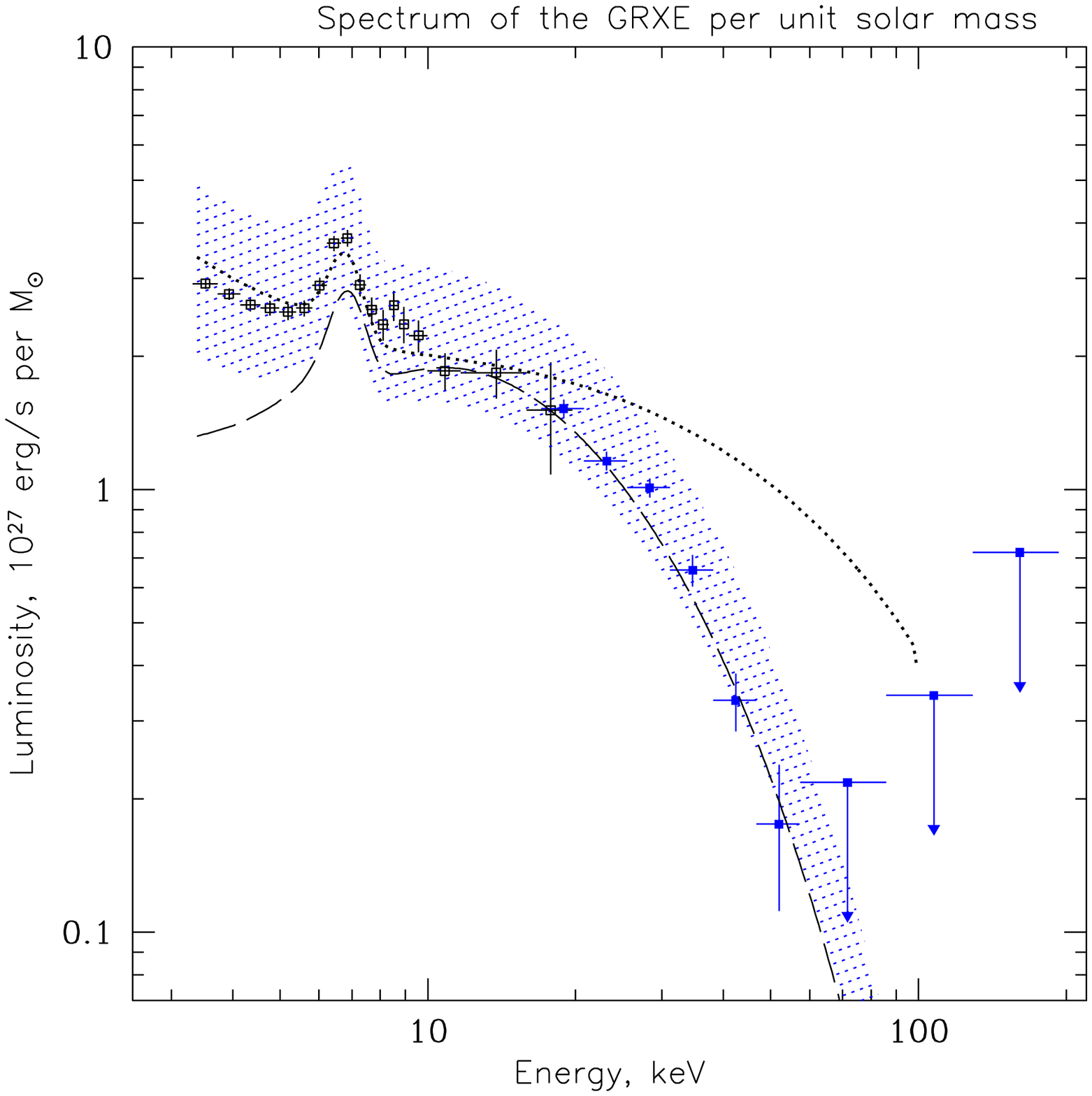}
\caption{Broad band spectrum of GRXE  per unit stellar mass. Blue
points represent result of this work. Shaded region represents a
"toy" composite spectrum of weak Galactic X-ray sources with weights
according to \cite{sazonov06}. For the input template spectrum of
intermediate polars we adopted a white dwarf mass $M_{\rm
wd}=0.5M_{\odot}$. The approximate (due to uncertainties in the relative
weights of CVs and coronally active stars) contribution of magnetic CVs
(intermediate polars and polars) to the GRXE emissivity is shown by the
dashed curve. An example of a composite GRXE spectrum with an adopted mass
of white dwarfs $M=1 M_\odot$ is shown by the dotted curve.}
\label{fig:spectrum}
\end{figure}

After we showed that hard GRXE volume emissivity traces the stellar
mass density in the Galaxy, we can construct a broad-band unit
stellar mass spectrum of the GRXE. We obtained ratios of hard X-ray
surface brightness in each energy band to NIR surface brightness.
NIR surface brightness was obtained after convolution of
reddening-corrected COBE/DIRBE $4.9\mu$m measurements with the IBIS
collimator response function in appropriate energy bins. Using the
obtained ratios and unit stellar mass 4.9$\mu$m luminosity we
calculated the unit stellar mass hard X-ray spectrum of the GRXE
(Fig.\ref{fig:spectrum}).

\section{Discussion}

\subsection{GRXE spectrum}

The obtained spectrum of the GRXE can now be compared with a composite
spectrum of known types of weak Galactic X-ray sources.
Unfortunately we do not have broad band spectra of all sources that
were used in the construction of the luminosity function of weak Galactic
X-ray sources \cite{sazonov06}. Therefore we tried to obtain some
"toy" composite spectrum that would posess the main properties of
the ideal sample of sources.

As the input templates of spectra of individual classes of sources we
take: the spectrum of V711 Tau as an RS CVn binary, AM Her as a polar, and
SU UMa as a dwarf nova. For the spectrum of intermediate polars, which are
the dominant contributors in hard X-rays, we adopt the model spectrum
of \cite{suleimanov05} with the white dwarf mass $M_{\rm wd}=0.5M_{\odot}$.

The important difference of this "toy" composite spectrum from that
used in the work of \cite{mikej05} is the value of the white dwarf
mass in the intermediate polar binary system. The temperature of the
optically thin plasma emitting X-ray radiation in the case of
accreting magnetic CVs (in particular - in intermediate polars,
which dominate in hard X-rays) strongly depends on the mass of the
white dwarf \citep[e.g.][]{aizu73}. In the range of masses $\sim
0.3-1.0M_\odot$ the optically thin plasma temperature, which is a
measure of the virial temperature of protons near the white dwarf,
is approximately $kT\propto M^{1.6-1.7}$ \cite[see e.g. WD
mass-radius relation in ][]{nauenberg72}. Therefore, it would be
more reasonable to use the average mass of the white dwarfs in the
Galaxy rather than some peculiar mass value. \cite{mikej05} used the
spectrum of the binary system V1223 Sgr, which harbors a white dwarf
with mass $M_{\rm wd}\sim 1 M_{\odot}$ \citep[e.g.][]{suleimanov05},
while the average mass of white dwarfs in the Galaxy is apparently
considerably smaller -- $M_{\rm wd}\sim0.5 M_{\odot}$
\cite[e.g.][]{bergeron92,bergeron95,kool92,politano96}. Therefore,
the hard X-ray part of the true spectrum of the GRXE if it were
composed of $\sim0.5M_{\odot}$ white dwarfs is expected to be
significantly softer than shown in Fig.8 of \cite{mikej05}. This is
indeed seen in Fig.\ref{fig:spectrum}.

Note than according to the above reasoning {\em the shape of the
GRXE spectrum in hard X-rays can be used to determine the average
mass of the white dwarfs in accreting magnetic CVs in the Galaxy}.
The exact determination of the average WD mass is subject to
uncertainties of relative contribution of different types of
Galactic X-ray sources, but our first estimate shows that it is
approximately consistent with $\langle M_{\rm wd,IP} \rangle\sim0.5
M_{\odot}$.

At the energies 60-200 keV we did not detect any hard X-ray emission
of the Galaxy apart from the contribution of a relatively small number of
bright point sources visible by IBIS/ISGRI.

At energies higher than 100-200 keV a more detailed study of the
IBIS/ISGRI detector background is needed in order to recover the
properties of the unresolved emission in the Galaxy. In addition to
the instrumental problems at these energies there is a strong
contribution of the diffuse positronium continuum in the Galactic
Center region, which should be carefully taken into account. We plan
to study the unresolved Galactic continuum at these high energies in
our future work.

\subsection{IGR~J17456-2901}

Images of the Galactic Center obtained by IBIS/ISGRI
\cite[e.g.][]{revnivtsev04,belanger05} showed a bright spot at the
position of the Galactic Center (Sgr A$^*$), which was designated as
an INTEGRAL source IGR~J17456-2901. Association of this source with
a low-mass X-ray binary located near  Sgr A$^*$ \citep{revnivtsev04}
or with emission from Sgr A$^*$ itself \cite[e.g.][]{belanger04}
does not allows one to explain the properties of the source. In
particular, it was shown that this source is most likely not a point
like source and does not consist of a small number of bright LMXBs
\citep{neronov05}. \cite{belanger05} argued that the emission of
IGR~J17456-2901 might have the same origin as the source of
ultra-high energy (TeV) photons seen by HESS \citep{aharonian06}.

\begin{figure}[htb]
\includegraphics[width=\columnwidth,bb=30 150 600 730,clip]{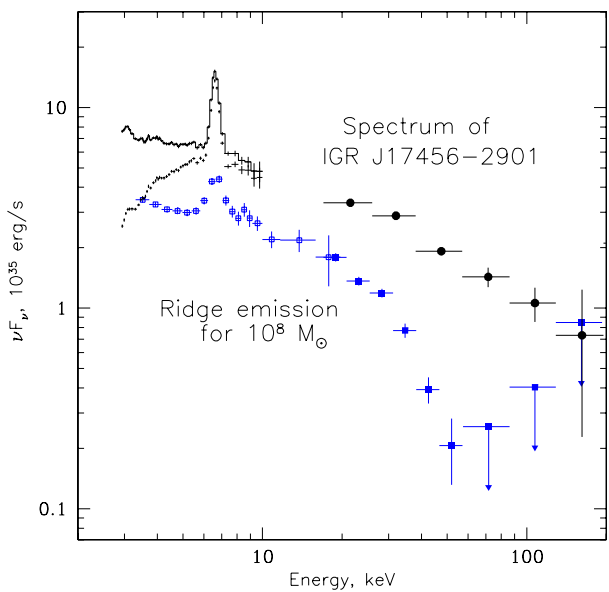}
\caption{The spectrum of the Galactic ridge X-ray emission
calculated for a $10^8 M_\odot$ stellar mass (blue squares). The spectrum
of Galactic Center source IGR~J17456-2901 is shown by black symbols.
Measurements by INTEGRAL/IBIS are shown by filled symbols. The hard
X-ray spectrum is complemented by a spectrum of integrated emission
of the central $10^\prime$ around Sgr A$^*$, measured by ASCA/GIS (lower
spectrum showed by black crosses) and corrected for interstellar
absorption $N_H=7\times10^{22}$ cm$^{-2}$ (upper spectrum showed by
black crosses).}\label{fig:central}
\end{figure}

It was argued by \cite{mikej05} based on the proportionality of the
GRXE volume emissivity to the Galactic stellar mass density that the
emission of IGR~J17456-2901 should consist (or contain a large
contribution) of the integrated (due to the limited angular
resolution of IBIS/ISGRI) emission of all the stars
($\sim10^{8}M_{\odot}$) within central $\sim30$pc ($\sim12^\prime$
at the Galactic Center distance, which is approximately equal to the
INTEGRAL/IBIS angular resolution) around Sgr A$^*$. Therefore, we
could anticipate that the spectrum of IGR~J17456-2901 would be
similar to that of the true GRXE \citep[e.g.][]{mikej05}. This is
partially true (see Fig.\ref{fig:central}).

The spectrum of the central $\sim10^\prime$ around Sgr A$^*$ is very
similar to that of the GRXE at least in the energy band $\sim 3-30$
keV, except for the somewhat higher normalization, which would
correspond better to an integrated stellar mass $\sim2 \times 10^{8}
M_\odot$ rather then $10^{8} M_\odot$ as we assumed. Whether this is
just due to the uncertain value of the total stellar mass within
$\sim30$ pc around Sgr A$^*$ \cite[e.g.][]{lindqvist92}, or in fact
due to a different value of X-ray emissivity per unit stellar mass
in this area \footnote{We would like to note that for the
near-infrared spectral band, the mass to light ratio was found to be
different within central 30 pc around Sgr A$^*$ and in the rest of
the Galaxy, see e.g. \citealt{launhardt02}} is a subject of a
special detailed study which is beyond the scope of this paper.

However, as can be seen from Fig. \ref{fig:central}, the spectrum of
IGR~J17456-2901 at energies 60-100 keV is obviously harder than that
of the GRXE.

Below we outline several possible origins of the observed spectral
difference.

{\it Confusion?}

The surroundings of the Galactic Center is a very complex region populated
by different types of compact and diffuse sources of emission.
The telescope IBIS, whose data were used here to derive the spectrum of
IGR~J17456-2901, has angular resolution $\sim12^\prime$,
therefore we cannot exclude that some emission region spatially
distinct from the nuclear stellar cluster contributes to the flux
that we measure from the position of IGR~J17456-2901.

This possibility finds relatively solid support from the fact that the
peak of hard X-ray emission at energies $>70-80$ keV, where the
contribution of compact sources visible at lower energies
vanishes, has a large offset with respect to Sgr A$^*$  -- $\sim
5.7^\prime$ \citep[see also][]{belanger05}. At the peak of the hard
X-ray emission there are no known persistently bright compact X-ray
sources \cite[see e.g.][]{sakano02}, while there has been observed high
energy (GeV and TeV) non-thermal emission, presumably originating as
a result of interaction of cosmic rays with interstellar matter
\citep{aharonian06}. If the spectrum of this non-thermal emission
has a relatively flat slope (e.g. $\Gamma\sim2$), then its relative
contribution in the standard X-ray energy band ($<10$ keV) should be very
small and should strongly rise with energy, therefore leading to the
observed difference in the spectra of IGR~J17456-2901 and the GRXE.

Another example of possible confusion of different sources is a
strongly photoabsorbed hard X-ray source, which is bright in hard
X-rays but practically absent in the standard X-ray band, located at
$6^\prime$ of Sgr A$^*$.

If, to the contrary, the hard X-ray emission of IGR~J17456-2901 does
belong to the Galactic nuclear stellar cluster, then different
possible scenarios may be proposed.

{\it Additional population of sources in the nuclear stellar
cluster?}

The nuclear stellar cluster is one of the densest regions in the Galaxy.
The stellar density within a few parsecs of Sgr A$^*$ exceeds $10^{5}$
stars pc$^{-3}$. It is not unreasonable to assume that such an exotic
environment may provide conditions for formation of a population of
X-ray emitting systems different from those in the rest of the Galaxy.
Tens or hundreds of sources per the $\sim 10^8 M_\odot$ stellar mass of
the central 30 pc with luminosities $\sim10^{32}$ erg/s and hard
X-ray spectra could significantly contribute to the observed
difference between the spectra of the GRXE and IGR~J17456-2901.

{\it Massive white dwarfs?}

If the average mass of accreting white dwarfs in the Galactic Center
region is systematically higher than in the rest of the Galaxy,
we can also anticipate a considerably harder cumulative spectrum of
weak X-ray sources. Induced massive white dwarf binaries formation
due to tidal capture process in dense environment, similar to
that in globular clusters  \cite[e.g.][]{ivanova06}, is not very
likely due to the much higher velocities of stars in the nuclear stellar
cluster. In principle, more massive white dwarfs might concentrate
in regions with deeper gravitational potential due to gravitational
mass segregation, similar to what is observed in globular clusters
\cite[e.g.][]{grindlay06}, but in the case of our Galactic Center
this scenario is also not very likely \cite[see e.g. simulations in
][]{freitag06}.

\section{Conclusion}

1) We have shown that the surface brightness distribution of the
GRXE in the energy band 17-60 keV very closely follows the near
infrared surface brightness distribution throughout the Galaxy. This
strongly supports the conclusion of \cite{mikej05} based on
lower energies (3--20 keV) data. The surface brightness
distributions of the gamma--ray background (EGRET data, 30 MeV-10
GeV), neutral interstellar matter ($HI$ map), and molecular
interstellar gas ($CO$ map) do not show such correspondence with
the hard GRXE intensity. The hard X-ray (17--60 keV) emissivity of
the Galactic ridge, recalculated per unit stellar mass is $(0.9-1.2)
\times10^{27}$ erg s$^{-1}$ M$^{-1}_{\odot}$. This value is in
good agreement (after correction for the energy band) with the unit
stellar mass X-ray emissivity of weak Galactic X-ray sources
\citep{sazonov06,mikej05}. The total Galactic hard X-ray luminosity
of the GRXE is $(4.2 \pm 0.3)\times10^{37}$ \lum in the 17--60~keV
energy band.

However, we should note that the difference in the morphology of the
EGRET Galactic gamma--ray background and hard X-ray (17-60 keV)
ridge emission observed by IBIS/ISGRI cannot by itself be considered
as a strong argument against the hypothesis of the cosmic ray origin
for the GRXE. Indeed, if the hard X-ray background emission of the
Galaxy were dominated by bremsstrahlung of low energy
($\la0.5$MeV) cosmic ray electrons \citep[see
e.g.][]{stecker77,mandrou80,sacher84,harris90}, then these electrons
might be confined to an almost immediate vicinity of their birthplace
if the interstellar magnetic field is sufficiently tangled. This would
happen because electrons at these energies have very small mean free
paths in the presence of tangled interstellar magnetic field
\citep[e.g.][]{zwickl78}. If the places of origin of such electrons
somehow followed the stellar mass distribution in the Galaxy, then
the hard X-ray background, induced by such cosmic--rays would also
follow the NIR intensity distribution.

2) Subtracting the flux of detected point sources from the total
IBIS aperture sky flux we have obtained the spectrum of the GRXE.
Its shape well agrees with the spectral shape of accreting magnetic
white dwarfs, which are expected to provide a dominant contribution to
the Galactic X-ray emission in this energy band. The shape of the spectrum of
the GRXE allows us to estimate the average mass of accreting
magnetic white dwarfs in the Galaxy $\langle M_{\rm wd} \rangle \sim
0.5 M_\odot$.

3) We have shown that the Galactic background emission is
undetectable in the energy range $\sim$60-200 keV. The signal that was
previously ascribed to the Galactic background emission at these
energies was most likely due to emission of unresolved point
sources.

\begin{figure*}[t]
\hbox{
  \includegraphics[width=\textwidth]{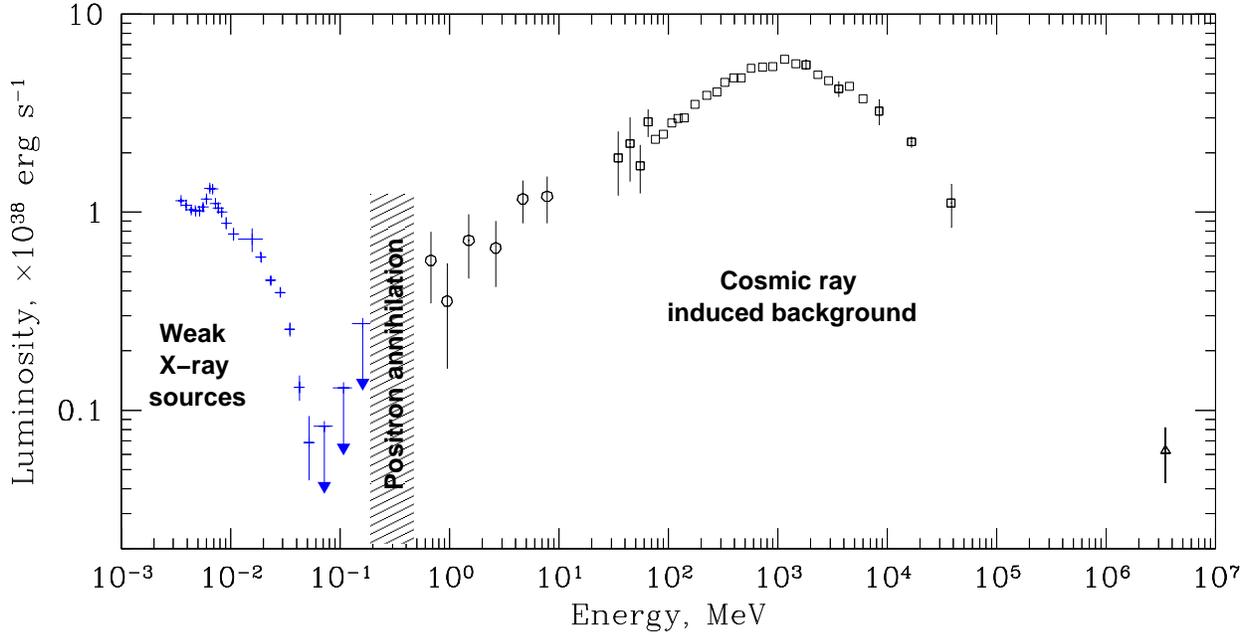}
}
  \caption{Schematic luminosity spectrum of "unresolved" emission of the Galaxy
  in the energy band 3 keV -- 4 TeV. In the X-ray energy band the
  luminosity spectrum was
scaled from the GRXE unit stellar mass emissivity spectrum (Fig.
\ref{fig:spectrum}) assuming a mass of the Galaxy of $3.9\times10^{10}
  M_\odot$.
For scaling the $\gamma$-ray part of the spectrum we adopted a value
of the total Galactic luminosity at $>100$ MeV of  $L_{>100 \rm
MeV}=2\times10^{39}$ erg/s \citep{bloemen84}. Measurements in
$\gamma$-rays by CGRO/OSSE and CGRO/EGRET are adopted from
\cite{kinzer99}, the measurement at TeV energies is rescaled from
\cite{atkins05,prodanovic06}. The shaded region at energies
$\sim$200-500 keV denotes the area where positron annihilation
radiation in the Galactic Center region strongly dominates}
\label{gal_emission_bkg}

\end{figure*}

4) Our results fit in the model in which the Galactic ridge X-ray
emission in energy band 3-100 keV originates as superposition of
weak Galactic point sources. This suggests that at energies $\ga$
200 keV a change of the nature of the unresolved Galactic emission
to cosmic--ray induced background should occur. In order to
illustrate this we present a scheme of the luminosity spectrum of
unresolved emission of the whole Galaxy in
Fig.\ref{gal_emission_bkg}. One should remember that according to
our model the ratio of the $\gamma$-ray to X-ray unresolved
background luminosities strongly varies across the Galaxy, therefore
the presented broad-band spectrum should be considered as only a
schematic representation of the real luminosity spectrum of the
Galaxy.

As the Galactic Center region at energies 200-500 keV contains a
powerful diffuse emission of the positronium continuum that is
very hard to disentangle from the cosmic-ray induced radiation we
anticipate that an answer to the question where the cosmic-ray
induced radiation begins to dominate can be obtained only either by
studying regions away from the Galactic Center or at energies 0.5-10 MeV.

\begin{acknowledgements}
This research was done thanks to unique capabilities of the INTEGRAL
observatory. The data used were obtained from the European and Russian
INTEGRAL Science Data Centers and from the High Energy Astrophysics
Science Archive Research Center Online Service of the NASA/Goddard
Space Flight Center. We acknowledge the use of the Legacy Archive for
Microwave Background Data Analysis (LAMBDA). Support for LAMBDA is
provided by the NASA Office of Space Science. Authors thank Hans
Ritter for useful discussion regarding WD masses in the Galaxy. The
work was supported by the President of the Russian Federation (through
the program of support of leading scientific schools, project
NSH-1100.2006.2), by the Presidium of the Russian Academy of
Sciences/RAS (the program ``Origin and evolution of stars and
galaxies''), by the Division of Physical Sciences of the RAS (the
program ``Extended objects in the Universe''), and by the Russian
Basic Research Foundation (the project 05-02-16540).

\end{acknowledgements}

\bibliography{ridge,integral}

\end{document}